\documentclass[aps,prb,twocolumn,10pt,showpacs,superscriptaddress]{revtex4-1}

\usepackage{amsmath}
\usepackage{graphicx}
\usepackage{times}
\usepackage{amssymb}
\usepackage{hyperref}
\usepackage{textcomp}
\usepackage{xcolor}
\usepackage[T1]{fontenc}

\begin{document}
%\showthe\linewidth
\title{Rydberg excitons in the presence of an ultralow-density electron-hole plasma}
\author{J. Heck\"otter}
\author{M. Freitag}
\author{D. Fr\"ohlich}
\author{M. A{\ss}mann}
\author{M. Bayer}
\affiliation{Experimentelle Physik 2, Technische Universit\"at Dortmund, 
D-44221 Dortmund, Germany}
\author{P. Gr\"unwald}
\author{F. Sch\"one}
\author{D. Semkat}
\author{H. Stolz}
\author{S. Scheel}
\affiliation{Institut f\"ur Physik, Universit\"at Rostock, 
D-18051 Rostock, Germany}

\begin{abstract}
We use two-color pump-probe spectroscopy to study Rydberg excitons in Cu$_2$O in the presence of free carriers injected by above-band-gap excitation. Already at plasma densities $\rho_\text{eh}$ below one hundredth electron-hole pair per \textmu m$^{3}$, the Rydberg exciton absorption lines are bleached while their energies remain constant, until they finally disappear, starting from the highest observed principal quantum number $n_\text{max}$.  As confirmed by calculations, the band gap is reduced by many-particle effects caused by free carriers scaling as $\rho_\text{eh}^{1/2}$. An exciton line looses oscillator strength when the band edge approaches the exciton energy vanishing completely at the crossing point. We quantitatively describe this {\sl plasma blockade} by introducing an effective Bohr radius that determines the energy distance to the shifted band edge.  In combination with the negligible associated decoherence this opens the possibility to control the Rydberg exciton absorption through the plasma-induced band gap modulation.

\end{abstract}

\maketitle

\textsl{Introduction.} 
Coulomb-correlated two-particle complexes in the presence of free charges are of great interest in areas such as atomic physics, plasma physics or condensed-matter physics and represent a central problem of interacting many-body physics. In solids, the most prominent bound two-particle complex is the exciton formed by a negatively charged electron and a positively charged hole. The interactions between excitons as well as of excitons with free carriers have been studied intensely by linear and non-linear spectroscopy, addressing mostly the ground state in prototype semiconductors such as GaAs bulk crystals and quantum wells.~\cite{Capizzi,Collet85,Klingshirn} Carrier densities are typically assessed relative to the Mott density,~\cite{Mott,Norris82} which is the density where excitons in their ground state dissociate into an electron-hole plasma and are no longer optically excitable. This density can be roughly estimated by the condition that the inter-carrier separation becomes equal to the exciton Bohr radius.

In linear spectroscopy of excitons, such as photoluminescence, hardly any changes of the emission line are observed up to densities coming close to the Mott density: only then a low energy shift combined with a line broadening can be identified.~\cite{Klingshirn} In non-linear spectroscopy, such as four-wave mixing, exciton-exciton scattering  was shown to lead to decoherence for densities at least an order of magnitude below the Mott density.~\cite{Klingshirn,Shah} Limited resolution and sensitivity combined with considerable homogeneous and/or inhomogeneous broadening have prevented studies of interaction effects at lower densities.

Here we address the interaction of excitons not in their ground state but in states of large principal quantum numbers with carriers at densities many orders of magnitude below the Mott density -- a problem that turns out to be intimately related to the question of the highest principal quantum number $n_\text{max}$ that can be observed. This high-resolution study is facilitated by the recent observation of such highly excited excitons in absorption spectroscopy with linewidths in the \textmu eV-range in cuprous oxide.~\cite{Kazimierczuk2014} Specifically, we consider Rydberg states up to $n>20$ in the hydrogen-like exciton series. From their huge extension in the \textmu m-range, giant interaction effects arise: when an exciton is excited into a Rydberg state, excitation of another exciton in close vicinity of the first one is prevented due to the shift of the absorption by the interaction energy between the excitonic dipoles -- an effect known in atomic physics as Rydberg blockade.~\cite{Lukin01} Estimates indicate a few \textmu m-sized blockade volume in which the exciton absorption is blocked.~\cite{Kazimierczuk2014} These studies suggest that Rydberg excitons provide unique possibilities to understand the exciton-plasma interaction at very low carrier densities.

In detail, we study the impact of an ultralow-density plasma with electron-hole-pair densities down to less than $\rho_{eh} =$ 0.01 \textmu m$^{-3 }$ on Rydberg excitons at $T$ = 1.35~K. When exciting free carriers above the band edge, already extremely low pump powers modify the exciton spectrum by reducing the absorption of the discrete spectral lines. On the other hand, their energies remain unchanged as do their spectral linewidths, indicating negligible decoherence induced by the free carriers. Ultimately, exciton lines disappear, starting from the highest observed $n_\text{max}$, as a result of the band gap reduction to to the plasma. The density of less than one hundredth of an electron-hole pair, at which the absorption of the \textmu m-sized excitons can be controlled by the plasma-induced band gap modulation, indicates a ``sub-quantum'' control by fractional charge within the exciton volume.

{\sl Experiment.} For the absorption studies, a thin Cu$_2$O crystal slab with a thickness of 30 \textmu m is 
placed in the variable temperature insert of a Helium cryostat. For excitation we use two frequency-stabilized dye lasers. One laser, the probe, is used to scan the exciton spectrum by 
tuning the emitted photon energy, while the other pump laser is used for excitation of electron-hole pairs by fixing its photon energy to a value above the band 
gap of cuprous oxide which is located at 2.172~eV. The pump spot size of 0.3~mm is 
chosen slightly larger than the probe spot size to ensure homogeneous excitation 
conditions. The output of the probe laser is stabilized by a noise-eater. After 
transmission through the sample, the probe laser light is detected by a 
photodiode.

\begin{figure}[t]
\includegraphics[width=1\columnwidth]{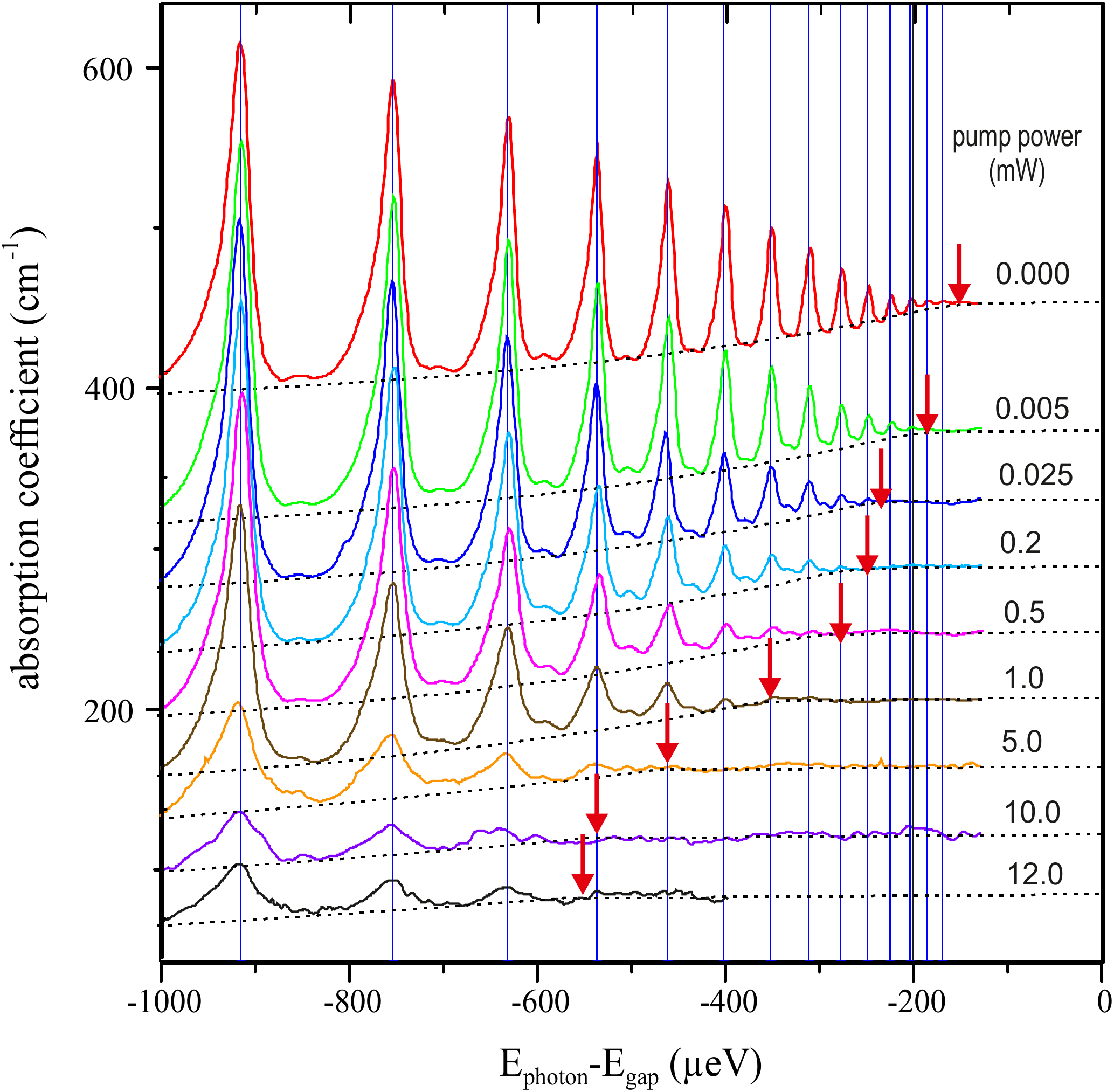}
\caption{Absorption spectra of Rydberg excitons from $n=10$ onwards recorded at $T=1.35$~K for simultaneous application of a pump laser with photon energy 
2.2~eV, operated at different powers. The top trace shows the absorption 
without pump. The dashed lines show the background absorption due to phonon-assisted processes plus the absorption into the continuum states. The arrows indicate the position of the band edge. Traces are shifted vertically for clarity.}
\label{fig.plasmaExp}
\end{figure}

Figure~\ref{fig.plasmaExp} shows absorption spectra of Rydberg excitons recorded at $T=1.35$~K. Without 
pump beam the series of P-exciton states can be observed up to $n=23$ (top trace). The decrease of 
the area below each absorption line with increasing principal quantum number shows some deviations 
from the expected scaling\cite{Elliott1957} of the exciton oscillator strength in an unpumped crystal:
while for $n \leq 15$ the observed peak area follows well the $n^{-3}$ scaling, for $n = 20$, e.g., it is reduced by a factor of 3 compared to the expectations as already observed in Ref.~\onlinecite{Kazimierczuk2014}, and this deviation increases further with $n$. 
The absorption lines appear on a continuously increasing background which should have two contributions. One is associated with optical 
phonon-assisted absorption into 1S yellow and green excitons.~\cite{ItohNarita} In the small energy range shown of 1~meV only, it is almost constant (variation < 1\%). The second one is due to absorption into continuum states by direct forbidden transitions, which should follow a step like behavior at the band edge.\cite{Elliott1957} Instead, the absorption shows a smooth rise towards the band edge (see the dashed line at the top trace in Fig.~1) which can be approximated well by an exponential tail similar to the well-known Urbach tail in the optical response of semiconductors.~\cite{urbachtail} Noteworthy, the transition to the constant behavior does not occur at the nominal band edge but is shifted to lower energies by about $150$~\textmu eV. We attribute both efffects to the influence of residual charged impurities due to non-complete compensation in Cu$_2$O \cite{ptype} that is typically p-type. These impurities provide on the one hand a disordered potential, giving rise to the Urbach tail, and on the other hand lead to screening.  

The pump photon energy is fixed at 2.20~eV, which is 28~meV above the band edge. 
The pump power is varied over more than three orders of magnitude from 5~\textmu W to 12~mW. 
Already for the lowest pump powers, the absorption becomes considerably reduced, 
leading to the disappearance of exciton resonances. This is associated with a further down shift of the apparent band edge position that can be accurately assessed by the transition from the Urbach tail to the flat continuum absorption (see the arrows in Fig. 1).  
%A best effort estimate assuming an electron-hole-pair lifetime of 18~ns and a recombination rate of $1.15\cdot 10^{-12}$cm$^3/$ns (see SOM)
%shows that the lowest power corresponds to a remarkably low plasma density $\rho_\text{eh}$ below 0.01 \textmu m$^{-3}$!
More and more lines vanish with increasing power starting from the highest ones, until for
the strongest applied power only the states up to $n=12$ are observed. 

Particularly noteworthy is that during this evolution the energies of the 
exciton lines remain constant within a fraction of their linewidth and, within the range where it can be reasonably estimated, also the 
linewidths of the resonances are unchanged. Therefore, the disappearance of 
the excitons cannot be related to scattering processes with the plasma because that would 
lead to lifetime reduction and resonance broadening. Rather it has to be related to a change of the exciton envelope functions due to the presence of the plasma. This {\it dressing} by a plasma is well known in atomic plasmas\cite{Kremp} but has never been considered in solid state physics up to now.

\begin{figure}[th]
\includegraphics[width=1\columnwidth]{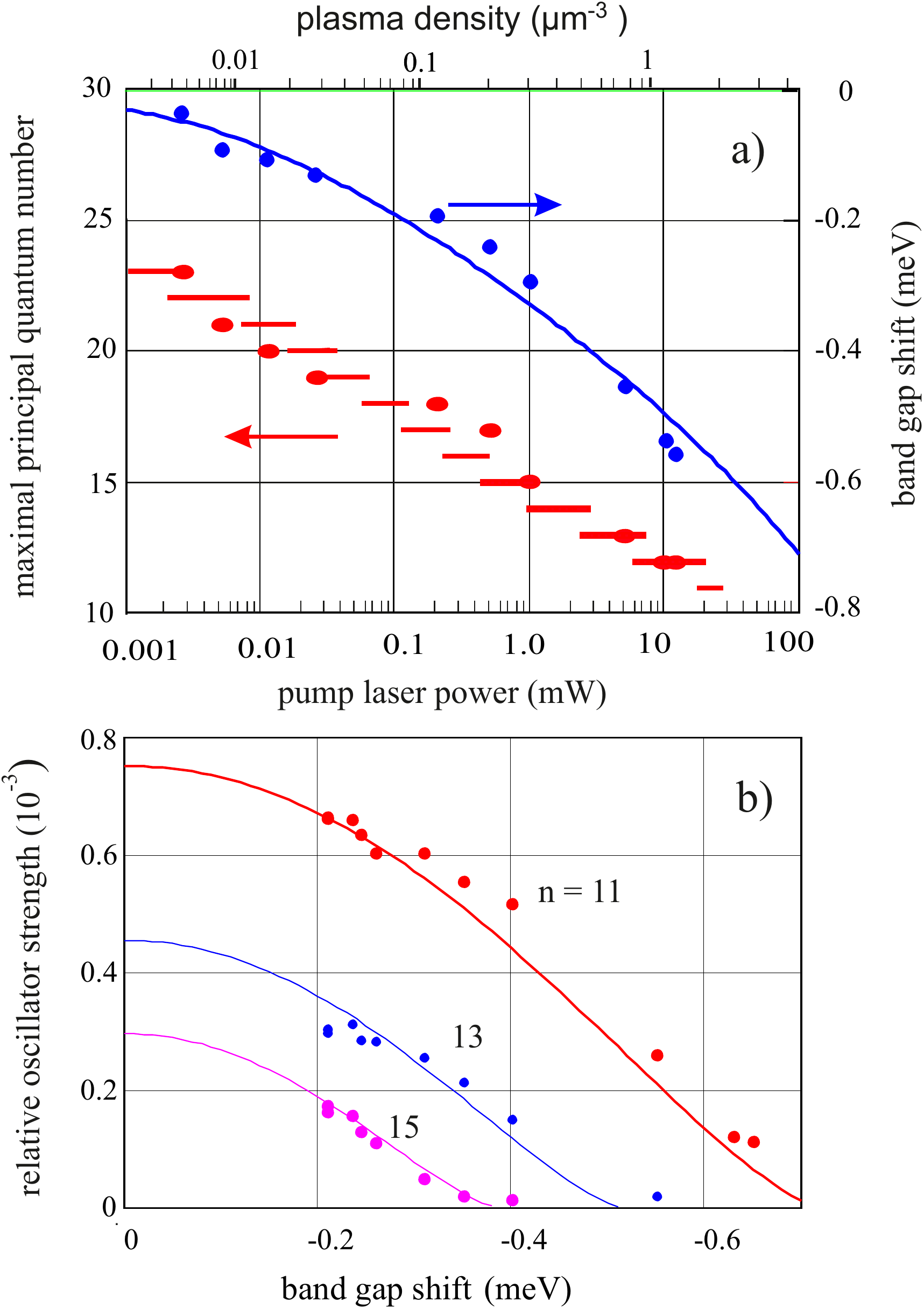}
\caption{%Maximal observable exciton state with $n_\text{max}$ {\sl vs} pump laser intensity for crystal temperatures of 1.3~K, 10~K, 22~K.
(a) Comparison of experiment (red dots) and theory (red bars) for the maximum observable exciton state $n_\text{max}$ vs.\ pump laser intensity for a crystal temperature $T=1.35$~K with $\hbar\omega_\text L=2.2$~eV (left ordinate) and dependence of band edge shift on pump laser power (right ordinate). To highlight the effect by the optically excited plasma, we subtracted from the experimental points (blue dots) the shift at zero power (see text). Blue line: band edge shift $\Delta_D$ as obtained from the Debye model using the densities obtained from the rate equations (see text).
(b)  Comparison of the experimental peak area\cite{footnote1} and the relative oscillator strength $f_{\mathrm{rel}}$ calculated from the effective Bohr radius and quantum number vs. pump laser power as discussed in the text for $n=11$ (red points and line), $n=13$ (blue) and $n=15$ (magenta).}\label{fig.result}
\end{figure}

Despite the different physical origin, the entire phenomenology is somewhat similar to the aforementioned Rydberg exciton blockade, so that the observations 
may be summarized by the term {\sl plasma blockade}. In Fig.~\ref{fig.result} the main experimental findings are summarized.  Panel (a) the variation of the band edge shift with pump power and panel (b) the decrease of the peak area of the absorption lines with the shift of the band edge (for $n_\text{max}$ as a function pump laser power see Fig.~3a).

{\sl{Theory.}}
From a theoretical point of view, the free carriers induced by above-band-gap excitation cause self-energy corrections 
and thus lower the band edge akin to the Mott effect.\cite{Zimmermann,Kremp,Semkat09} 
In $\mathrm{Cu_2O}$, the necessary plasma 
density\cite{manzke} for the fundamental 1S exciton to vanish into the continuum and being no longer excitable, is roughly $\rho_\text{Mott}\approx3 \cdot 10^{18}$~cm$^{-3}$. However, we are interested in the Mott-effect analogue for large values of $n$, 
that occurs at much lower plasma densities and hence laser intensities. 

The plasma degeneracy parameter is given by the product of the volume associated with the thermal wavelength $\Lambda_{th}(=\hbar\sqrt{2\pi/\mu_{eh}k_B T})$ times the plasma density, $\Lambda_{th}^3 \rho_{eh}$.~\cite{Kremp} In the low-density regime, the degeneracy parameter is much smaller than 1 as is the case in our experiments where even for large $\rho_{eh}=1$ \textmu m$^{-3}$ and a plasma temperature of 5~K (see Fig. 2b) it is about 0.15, so that the non-degenerate classical Debye theory can be applied for the plasma.~\cite{Kremp} Only for much smaller $T$ in the mK range the more general description established by Zimmermann \textit{et al.}\cite{Zimmermann78,Zimmermann} (for more recent elaborate investigations see Refs.~\onlinecite{Semkat09,Manzke12}) would be necessary. In the Debye model, the correlation part of the chemical potential $\Delta_D$ is a good measure for the band edge shift and depends on plasma density and temperature through $\Delta_D=-\kappa e^2/(4\pi\epsilon_0\epsilon_{\mathrm{b}})$ with $\kappa$ being the inverse screening length, which for a plasma with two components differing in effective temperature is given by
$\kappa^2=2\rho_\text{eh}e^2/(\epsilon_0\epsilon_{\mathrm{b}}k_{\mathrm{B}}T_{sc})$ with the effective screening temperature $1/T_\text{sc}=(1/T_\text{e}+1/T_\text{h})/2$. $T_{e}$ and $T_{h}$ are the plasma temperatures of electrons and holes, which in nonequilibrium situations differ from the crystal temperature $T_K$ as they are determined by relaxation and cooling of the carriers (see SOM).

\begin{figure}[th]
\includegraphics[width=1\columnwidth]{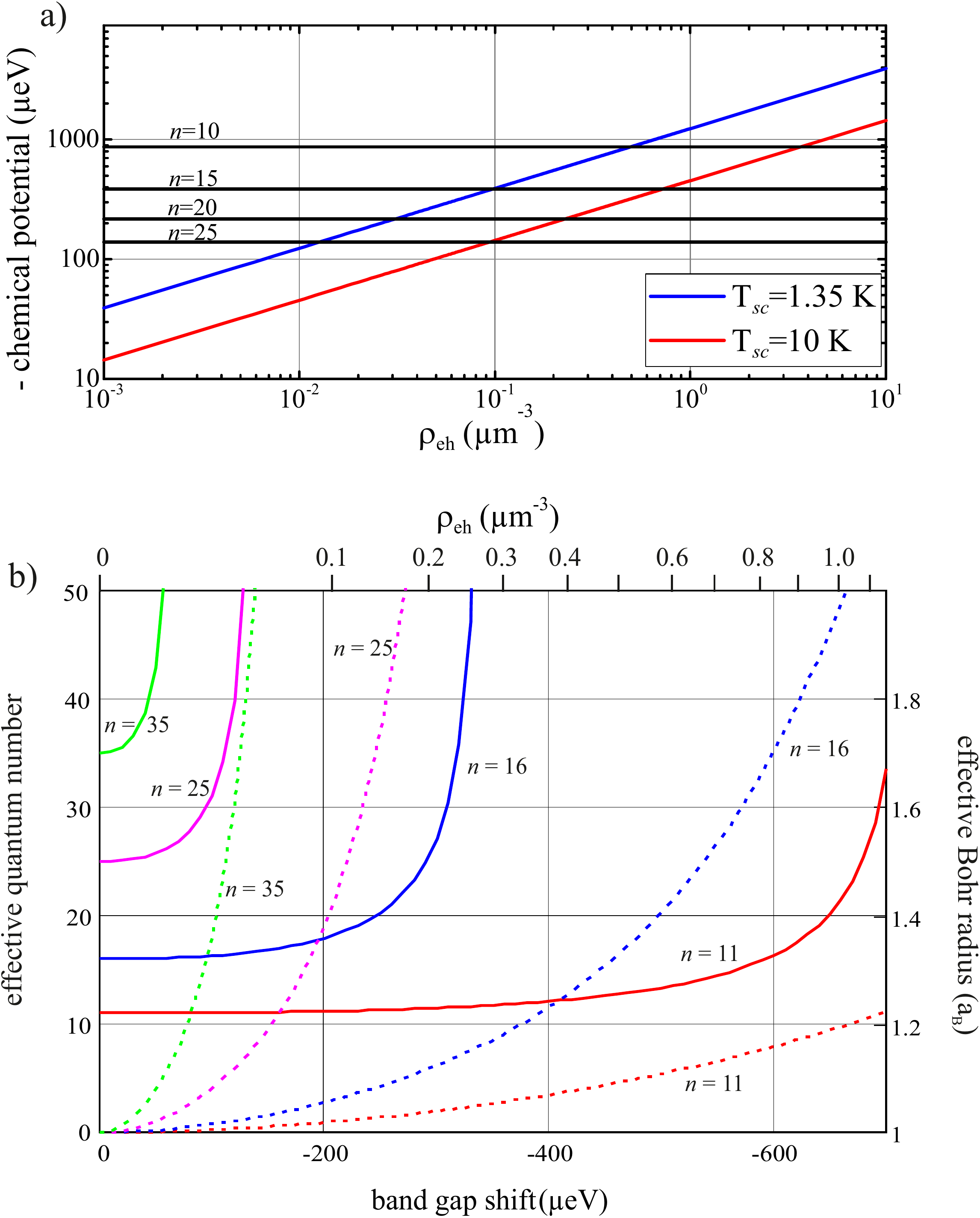}
\caption{Part a): Debye shift $\Delta_D$ vs.\ plasma density, at $T_{sc}=1.35$~K (blue line) and at $T_{sc}=10$~K (red line), whereby these values set the limits of the effective screening plasma temperatures (see Fig.~2b). The horizontal lines give the binding energies for the Rydberg excitons with $n=25,20,15,10$ (from bottom to top). 
Part b): Dependence of the effective Bohr radius $a_{\mathrm{Beff}}$ (in units of the Bohr radius $a_\text{B}$ (dashed lines, right ordinate) and the effective main quantum number $n_\text{eff}$ (full lines, left ordinate) on the band edge shift $\Delta$ for various main quantum numbers from 12 to 35. The corresponding density of the electron-hole plasma ($T_\text{sc}=5$~K) is shown as the upper abscissa.}\label{fig.DHZcomp}
\end{figure}
We then expect the exciton states to vanish when the chemical potential (see Fig.~\ref{fig.DHZcomp}a) is large enough that the band edge gets below the exciton state. For the $n=25$ exciton, the highest principal quantum number revealed so far in experiment,~\cite{Kazimierczuk2014} this happens already for densities as low as $\rho_\text{eh}\approx10^{-2}~\text{\textmu{}m}^{-3}$. 
For a density about two orders of magnitude larger, but still six orders below the 1S Mott density, the $n=10$ exciton can no longer be observed. From the Mott criterion we find for the dependence of the critical plasma density $\rho_\text{eh,c}(n)$, at which an exciton state $n$ vanishes into the continuum, a scaling with $n^{-4}$ which is in reasonable agreement with the experiments (see Fig.~2a).

%To appreciate these extremely low densities, we compare $\rho_\text{eh,c}(n)$ with the volume of the excitonic wave function $V_n=\tfrac{4}{3}\pi r_n^3 \propto n^6$, with $r_n$ being the Bohr radius of the respective state. We find $	\rho_\text{eh,c}(n)\cdot V_n\approx 10^{-3} n^2$. 
%This means that the $n=10$ exciton with an radius of about 165~nm is destroyed by only one unbound electron-hole pair in a volume ten times larger than its wave function, and underlines that Rydberg excitons are extremely sensitive monitors of charge fluctuations in their environment.

To be more quantitative, we have to connect the plasma density $\rho_\text{eh}$ to the applied laser power. The density obeys a rate equation $d\rho_\text{eh}/dt=G-\Gamma_\text{eh} \rho_\text{eh}-\Gamma_\text{rc} \rho_\text{eh}^2$
with the generation rate $G$ proportional to the pump power, the inverse lifetime $\Gamma_\text{eh}$, and the  bimolecular recombination rate $\Gamma_\text{rc}$ (for details, see SOM). The final plasma temperature also depends on the relaxation rates. The shorter the lifetime, the hotter is the plasma.
Thus the dependence of density and temperature (and, therefore, of the band edge shift) on the pump power is determined (see Fig.~2b) in effect only by the relaxation parameters of the plasma. To highlight the plasma influence, we have corrected the measured $\Delta$ by that due to impurities ($\Delta_0$ at $P_L=0$) by taking $\sqrt{\Delta^2-\Delta_0^2}$.  Choosing the appropriate parameters ($1/\Gamma_{eh}=18$~ns and $\Gamma_{rc}=1.15\cdot10^{-12}$~cm$^{3}$/ns), which can be done quite uniquely, results in the blue line in Fig.~2b. From the relaxation parameters, the dependence of the plasma density on laser power (see the bottom and top abscissa in Fig.~2a) can be obtained straightforwardly.
%Fortunately, we have determined the band edge shift directly from the spectra, so that we can derive the plasma influence without knowing $T_\text{e,h}$ and $\rho_\text{eh}$ in detail. From Fig.2b $\Delta$ decreases with laser power. Since $T_{e,h}$ will increase with power, the plasma density $\rho_{eh}$ must also increase. To obtain the exact dependence, one can vary the lifetime parameters and use the appropriate value for $T_{e,h}$. The solution as shown by the blue line turns out to be rather unique. The parameters found are . 
%The effective plasma screening temperature $T_\text{sc}$ is plotted as the dashed line in Fig.~2b. 

On this basis, we can analyze in detail the impact of the presence of a plasma on the exciton spectrum. Note that the reduction of the band gap is a first-order plasma process, while, in the
corresponding shift of the exciton resonances, various many-particle effects compensate each other to a large extent, e.~g. the self-energy, Pauli blocking, and screening of the interaction.~\cite{Zimmermann78,Kremp}
The calculations replicate the observed downward trend of $n_\text{max}$ quite well (see the theoretical dependence in Fig.~2a).
Similarly striking, however, is the decrease of oscillator strength of the lines starting already below the critical plasma density $\rho_\text{eh,c}(n)$ (see Fig. 2c). For $P$-excitons, the oscillator strength is determined by the square of the slope of the wave function at $r=0$, $f_{\mathrm{osc}}(n)\propto (\partial\phi(r)/\partial r|_{r=0})^2=(n^2-1)/(3 \pi a_\text{B}^5 n^5)$.\cite{Elliott1957} 
From related studies of hydrogen in atomic plasmas \cite{Kremp,Seidel,Paul} it is known that the wavefunctions are modified in two ways: On the one hand, they obey a modified Wannier equation (i.e., the Schr\"odinger equation for the relative motion of electron and hole) with the screened Coulomb  potential, that in a first approximation can be taken as the Debye potential $V_{\mathrm{D}}(r)=e^2/(4\pi\epsilon_0\epsilon_{\mathrm{b}}r)\exp(-\kappa r)$ that includes the Debye shift accounting for the band gap renormalization.~\cite{Kremp} A simple, but for low plasma densities quite accurate method, is to apply a variational procedure by adjusting the Bohr radius of the otherwise unchanged hydrogen wave function.~\cite{Paul} This gives the following expression for the energy of a P state with quantum number $n$ (with $x=a_\text{Beff}/a_\text{B}$ and $\xi= \kappa a_\text{B}/2$
\begin{eqnarray}
&&\langle E_\text{P}(n)\rangle=\Delta_D + \frac{Ry}{n^2}\\
&&\times\left(\frac{1}{x^2}- \frac{2}{x}\frac{1}{(1+n \xi x)^{2n}} \,_{2\!}F_1(-1-n,2-n,1,(n \xi x)^2)\right).\nonumber
\end{eqnarray}
Here $_2F_1$ is the Gaussian hypergeometric function.
The effective Bohr radii $a_{\mathrm{Beff}}>a_{\mathrm{B}}$ obtained in this way for different $n$ are shown in Fig.~3b. On the other hand, it is well known \cite{Kremp} that for higher plasma densities the simple Debye screening underestimates the renormalization of the wave function leading to a red shift of the energies and to an overestimation of the Mott densities. Only if the full dynamic screening is taken into account, resulting in an admixture of states with higher quantum numbers,\cite{Seidel} the correct behaviour of the states is obtained. In a qualitative way this can be taken into account by ascribing an effective quantum number $n_\text{eff}$ to the states so that the energy of the renormalized state $\langle E_\text{P}(n)\rangle$ stays equal to that of the unperturbed state independent of the band edge shift $\Delta_D$. The values of $n_\text{eff}$ obtained in this way are also shown in Fig.~3b.\cite{footnote2}
%we rather apply a picture which allows us to grasp the essence of the plasma effect on the exciton states and explain the observations. Looking at their energies relative to the plasma-shifted band edge, the effective ionization energy of the excitons $R_X/n^2+\Delta_D$decreases. Thus, an exciton with quantum number $n$ comes closer to the band edge, corresponding in effect to an exciton with larger Bohr radius  defined by
%\begin{equation}
%E(n_{\mathrm{eff}})=-\frac{R_X}{n^2}+\Delta_D\equiv -\frac{\hbar^2}{2\mu n^2 %a_{\mathrm{Beff}}^2}.
%\end{equation}
Starting from $a_\text{B}$, $a_{\mathrm{Beff}}$ increases already at relatively small band edge shifts and thus low densities (see the upper abscissa in Fig.~3b). On the other hand, $n_\text{eff}$ stays for low densities almost constant, but diverges then quite rapidly approaching the Mott density. 

Assuming that the dependence of the oscillator strength with Bohr radius and quantum number is also valid for the perturbed states, but now with $a_{\mathrm{Beff}}$ and $n_\text{eff}$, the oscillator strength and, therefore, the peak area depend via the band edge shift on density and temperature of the electron-hole plasma. A comparison of the relative oscillator strength ($f_{\mathrm{rel}}(n)= (n_\text{eff}^2-1)/((a_\text{Beff}/a_\text{B})^5 n_\text{eff}^5)$) obtained in this way with the peak area of the $P$ lines (Fig.~2c) shows a surprisingly good agreement, especially in view that there are no adjustable parameters, which convincingly demonstrates the applicability of our model. The description appears to be even correct at zero pump power, where the band edge shift is induced by uncompensated impurities. Therefore, we conclude that the highest observable principal quantum number is determined by the amount of uncompensated impurities which depends on temperature and can be directly read off from the effective band edge of the crystal. Indeed in the record breaking $n=25$ observations the effective band edge was shifted by only 110~\textmu eV. So to be able to observe higher $n$, one must be able to control this shift, e.g. by further lowering the crystal temperature into the mK range.  The best way, however, would be to have even purer samples available, which will be a great challenge for crystal growth.

{\textsl{Conclusions.}} We have studied the Rydberg exciton absorption under the influence of a driven electron-hole plasma. Inside a $\mathrm{Cu_2O}$ crystal, the plasma shifts the band edge downwards, diminishing the maximum excitable Rydberg state. Even without pumping, a band edge shift is present due to unavoidably uncompensated impurities which allows one to predict the necessary conditions to observe even higher Rydberg states than observed thus far.
Based on our understanding of the plasma influence on Rydberg excitons, we are able to control said influence, e.g., to precisely limit the maximum observable principal quantum number by injecting a single or very few electrons in the target volume. That leads to a fast modulation of this plasma blockade induced by the band gap modulation and could be an essential tool to manipulate Rydberg states in future technologies, even at the quantum level, without introducing any notable decoherence.

%\textsl{Acknowledgments.} 
\begin{acknowledgments}
We gratefully acknowledge support by the Priority Programme 1929, the SFB 652/3, and the TRR 160, all funded by the Deutsche Forschungsgemeinschaft.
\end{acknowledgments}


\begin{thebibliography}{99}

\bibitem{Capizzi} M. Capizzi, S. Modesti, A. Frova, J. L. Staehli, M. Guzzi, and R. A. Logan, Phys. Rev. B \textbf{29}, 2028 (1984).

\bibitem{Collet85} J. Collet,  
J. Phys. Chem. Solids \textbf{46}, 417 (1985). 

\bibitem{Klingshirn} C. F. Klingshirn, \textsl{Semiconductor Optics} (Springer, Berlin, 2012), and references therein.

\bibitem{Mott} N. F. Mott, 
Rev. Mod. Phys. \textbf{40}, 677 (1968).

\bibitem{Norris82} G. B. Norris and K. K. Bajaj,  Phys. Rev. B \textbf{26}, 6706 (1982).

\bibitem{Shah}  J. Shah, \textsl{Ultrafast Spectroscopy of Semiconductors and Semiconductor Nanostructures} (Springer, New York, 1999).

\bibitem{Kazimierczuk2014} T.~Kazimierczuk, D.~Fr\"ohlich, S.~Scheel, 
H.~Stolz, and M.~Bayer, 
 
Nature \textbf{514}, 343 (2014).

\bibitem{Lukin01} M. D. Lukin \textsl{et al.}, 
Phys. Rev. Lett. \textbf{87}, 037901 (2001).

% \bibitem{SchaeferWegener} W. Sch\"afer and M. Wegener, \textsl{
% Semiconductor Optics and Transport Phenomena}
% (Springer,Berlin,2002).

% \bibitem{Elliott61} R. J. Elliott,  
% Phys. Rev. \textbf{124}, 340--345 (1961).

\bibitem{Elliott1957}R. J. Elliott, 
Phys. Rev. \textbf{108}, 1384--1389 (1957).

\bibitem{ItohNarita} T. Itoh and S. Narita, 
J. Phys. Soc. Jpn. \textbf{39}, 140 (1975).

\bibitem{urbachtail} S. John, C. Soukoulis, M.H. Cohen, and E. N. Economou, 
Phys. Rev. Lett. \textbf{57}, 1777--1780 (1986).

\bibitem{ptype} H. J. Park and K. Natesan, Oxydation of Metals \textbf{39}, 411 (1993).

\bibitem{Kremp} D. Kremp, M. Schlanges, and W.-D. Kraeft,
\textsl{Quantum Statistics of Nonideal Plasmas} (Springer, Berlin, 2005).

% \bibitem{Florian16b} F. Sch\"one, S. O. Kr\"uger, P. Gr\"unwald, H. Stolz, S. 
% Scheel, M. A\ss{}mann, J. Heck\"otter, J. Thewes, D. Fr\"ohlich, and M. Bayer,
% J. Phys. B \textbf{49}, 134003 (2016).

\bibitem{Zimmermann} R. Zimmermann, Phys. Status Solidi B \textbf{146}, 371 (1988).

\bibitem{Semkat09} D. Semkat, F. Richter, D. Kremp, G. Manzke, W.-D. Kraeft, 
and K. Henneberger, Phys. Rev. B \textbf{80}, 155201 (2009).

\bibitem{manzke} G. Manzke, D. Semkat, F. Richter, D. Kremp, and K. Henneberger,  J. Phys.: Conf. Ser. \textbf{210}, 012020 (2010).

\bibitem{Zimmermann78} R. Zimmermann, K. Kilimann, W. D. Kraeft, D. Kremp, and G. R\"opke,
Phys. Status Solidi B \textbf{90}, 175 (1978).

\bibitem{Manzke12} G. Manzke, D. Semkat, and H. Stolz, New J. Phys. \textbf{14}, 095002 (2012).

\bibitem{footnote1} The peak area is normalized by a factor $0.78\cdot 10^{-8}$cm/\textmu eV which is obtained from comparing peak area and oscillator strength at lower $n$ not influenced by the plasma.

%\bibitem{Stolz12a} H. Stolz, R. Schwartz, F. Kieseling, S. Som, M. Kaupsch, S. Sobkowiak, D. Semkat, N. Naka, Th. Koch, and H. Fehske, New J. Phys. \textbf{14}, 105007 (2012).

%\bibitem{Stolz12b} R. Schwartz, N. Naka, F. Kieseling, and H. Stolz, New J. Phys. \textbf{14}, 
%023054 (2012).

%\bibitem{ridley5th} B. K. Ridley, \textsl{Quantum Processes in Semiconductors} (Clarendon Press, Oxford, 1999).

\bibitem{Paul}S. Paul and Y. K. Ho, Phys. Plasmas \textbf{16}, 063302 (2009).

% \bibitem{Gruenwald2016} P. Gr\"unwald, M. A\ss{}mann, J. Heck\"otter, D. 
% Fr\"ohlich, M. Bayer, H. Stolz, and S. Scheel, 
% Phys. Rev. Lett. \textbf{117}, 133003 (2016).
% 
% \bibitem{CardonaBook} P. Y. Yu and M. Cardona, 
% \textsl{Fundamentals of Semiconductors} (Springer, Berlin, 2005).

\bibitem{Seidel}J. Seidel, S. Arndt, and W. D. Kraeft, Phys. Rev. E \textbf{52}, 5387 (1995).
\bibitem{footnote2} Doing the extremalization of $a_\text{B}$ and $n$ simultaneously results essentially in the same dependences.
\end{thebibliography}
\end{document}